\documentstyle[12pt]{article}
\begin{document}
\baselineskip=15pt

\newcommand{\be}{\begin{equation}}
\newcommand{\ee}{\end{equation}}
\newcommand{\bq}{\begin{eqnarray}}
\newcommand{\eq}{\end{eqnarray}}
\newcommand{\x}{{\bf x}}
\newcommand{\p}{\varphi}
\newcommand{\Sc}{Schr\"odinger\,}
\newcommand{\del}{\nabla}
\newcommand{\A}{{\bf A}}
\newcommand{\nn}{\nonumber\\}

\begin{titlepage}
\rightline{DTP 95/47}
\vskip1in
\begin{center}
{\large Reconstructing the Vacuum Functional of Yang-Mills from its
Large Distance Behaviour}
\end{center}
\vskip1in
\begin{center}
{\large
Paul Mansfield

Department of Mathematical Sciences

University of Durham

South Road

Durham, DH1 3LE, England}

{\it P.R.W.Mansfield@durham.ac.uk}
\end{center}
\vskip1in
\begin{abstract}
\noindent
For fields that vary slowly on the scale of the lightest mass
the logarithm of the vacuum functional can be expanded as a sum of
local functionals. For Yang-Mills theory the leading term
in the expansion dominates large distance effects and leads to
an area law for the Wilson loop. However, this expansion cannot be
expected to converge for fields that vary more rapidly.
By studying the analyticity of the vacuum functional under scale
transformations we show how to re-sum this series so as to
reconstruct the vacuum functional for arbitrary fields.
\end{abstract}

\end{titlepage}

\section{\bf Introduction}

When the \Sc representation vacuum functional of  a massive quantum
field theory is evaluated for slowly
varying fields its logarithm reduces to a sum of local integrals.
For example, the vacuum functional of a free scalar field theory in
$D+1$ dimensions with mass $m$ is  $\Psi [\p ]\equiv\langle \p
|0\rangle=exp\, -{1\over 2}\int d^D\x\,
\p\sqrt{-\nabla^2+m^2}\,\p$, so that if the Fourier transform of
$\p$ vanishes
for momenta with magnitude greater than the mass, $m$, the logarithm
of $\Psi$ can be expanded in the convergent series
\be
ln\,\Psi=
-\int d^D\x \left({m\over 2}\p^2+{1\over 4m}(\nabla \p)^2-
{1\over 16 m^3}
(\nabla^2\p)^2..\right).\label{eq:one}
\ee
The terms of this expansion are local
in the sense that they involve the field and a finite number of
its derivatives at the same spatial point. The same is true for an
interacting
theory in which the lightest particle has non-zero mass,
because the logarithm of the vacuum functional is a sum of
connected Feynman diagrams, and since massive propagators are
exponentially
damped at large distances these diagrams reduce to local integrals
for slowly varying fields. In the case of 3+1 dimensional Yang-Mills
theory
the lightest glueball mass must depend non-perturbatively on the
coupling so
that the local expansion of the vacuum functional could not emerge
from the usual semi-classical expansion of the functional integral.
Nonetheless we would expect to obtain for slowly varying
gauge-potential
\bq W[{\bf A}]\equiv ln\,\Psi [{\bf A}]&=&\int d^3{\bf x}\,
(a_1tr\,{\bf B}\cdot{\bf B}/m +a_2tr\,D\wedge{\bf B}\cdot
D\wedge{\bf B}/m^3+
\nonumber\\
& &
a_3tr\,{\bf B}\cdot
({\bf B}\wedge{\bf B})/m^3 +a_4 tr\,{\bf B}\cdot{\bf B}
\,{\bf B}\cdot{\bf B}/m^5+..)
\label{eq:YMexp}
\eq
where $\bf B$ is the Yang-Mills magnetic field $\nabla
\wedge{\bf A}+{\bf A}\wedge {\bf A}$, and $D$ is the gauge
covariant
derivative. (This is qualitatitively different to the case of
pure Abelian
gauge theory where $W=-{1\over 4\pi e^2}\int d^3 {\bf x}\,
d^3{\bf y }\,
{\bf B}({\bf x})\cdot{\bf B}({\bf y })/({\bf x}-{\bf y})^2$ is
conformally invariant and cannot be expanded as an integral of
local quantities.)
The unknown coefficents $a_i$ are dimensionless
constants. In principle these coefficents can be determined from the
\Sc
equation. In \cite{Paul} we showed that the \Sc equation does not
take
its usual form because the process of removing the cut-off does not
commute
with expanding in terms of local quantities. However, by re-summing
in
terms of the cut-off an appropriate \Sc equation can be constructed,
and we
suggested a method of solution that does not rely on a semi-classical
perturbation theory. Alternatively the first few terms can be computed
in lattice gauge therory. Approximating the vacuum functional by the
first
term of (\ref{eq:YMexp}) was originally suggested by Greensite
\cite{Jef1}, who
used Monte Carlo techniques to compute its coefficent for the
gauge group $SU(2)$ in 2+1 dimensions
\cite{Jef2} and, with Iwasaki, in 3+1 dimensions \cite{Jef3}.
When these results are translated from lattice to continuum fields
they give $a_1/m={a\over 8}exp \,[{3\over 11}\pi^2(\beta-\beta_0)]$,
\cite{Jef3},
where $a$ is the lattice spacing, $\beta$ the conventional coupling
that enters the Wilson action for the 3+1 dimensional theory,
and $\beta_0\approx 1.74$. The 2+1 dimensional result was extended
by Arisue in
\cite{ari}, and the case of gauge group $SU(3)$
considered by Chen {\it et al} in \cite{chen}.
For $SU(2)$ the 2+1 dimensional theory yields
$W_3[\A ]=\int d^2{\bf x}\,\left({b_1\over e^4}\,tr\,B^2+
{b_2\over  e^8}\,
tr\,D_iB\,D_iB +..\right)$, \cite{ari}, where
$b_1= 0.91\pm 0.02$, $b_2=-0.19\pm 0.05$ and $e$ is the continuum
coupling constant
in 2+1 dimensions corresponding to the normalisation of the Lie
algebra generators such that $tr\,T^AT^B=-\delta^{AB}/2$.
(Note, that whilst the naive
continuum limit of the lattice strong coupling expansion
\cite{Kog} would yield a
similar expansion the coefficents would be wrong as they do not scale
appropriately in the continuum limit.)
To get some idea of the relevance of these results to large distance
effects we can attempt to compute the string tension. We will simplify
the
calculation by retaining only the leading term. In the 2+1 dimensional
theory
this leads to an area law for the Wilson loop \cite{halp} because the
leading
term in $W_3$ is the action for two-dimensional Yang-Mills which
is free and confining. Applying the  calculation given in \cite{Paul2}
to the result of \cite{ari} gives the string tension as $\sigma_3=
{3e^4/( 16 b_1)}
\approx 0.21 \,e^4$ which should be compared with
the recent direct Monte Carlo estimate $\sigma=(0.112\pm 0.001)\,e^4$
\cite{teper}.
Similarly the large Wilson loops in the 3+1 dimensional theory
may be estimated using just the first term in (\ref{eq:YMexp})
which reduces to a calculation in the three-dimensional theory,
and thence to a calculation in the two-dimensional theory,
so that the area law emerges from a kind of dimensional reduction
\cite{Olesen}. Putting
this together gives the string tension in the 3+1 dimensional
theory as $\sigma_4\approx 0.21\tilde e^4$ with $\tilde e^2=m/(2a_1)$
so that $\sigma_4 a^2\approx 3.4\,exp\,[-{6\pi^2\over 11}
(\beta-1.74)]$.
For $\beta=2.85$ this yields $\sigma_4 a^2\approx 0.0086$ which
compares
with the accurate direct Monte Carlo measurement $\sigma_4 a^2
=0.00363(17) $ \cite{UK}. Some comments are in order. Firstly, the
coefficents
$a_1,b_1,b_2$ are obtained by applying the naive continuum limit
to a fit of the lattice vacuum functional obtained for slowly
varying plaquette variables. This is justified since the transition
from
plaquette variables to continuum ones is itself a derivative expansion
in
continuum variables. If we had computed Wilson loops using the
form of the vacuum functional originally found in terms of plaquette
variables we would also have obtained an area law, but with a different
string tension that would not scale appropriately in the continuum
limit.
This is because the integration over plaquette variables would include
rapidly varying fields that were excluded when we made the transition to
continuum variables, and for which the form of the vacuum functional used
is inappropriate. The calculation of the string tension using just
$a_1$ and $b_1$, although numerically incorrect by about a factor of two,
does at least scale properly. Presumably the numerical value would be
improved
were we to include the other terms in the expansion of $W$.
In spite of this qualified
success in describing large Wilson loops we can only expect the
local expansion
to converge for configurations that vary slowly on the
scale of the mass of the lightest glueball so it would appear
not to be relevant, for example, to the computation of
the glueball spectrum as this involves heavier particles.
The purpose of this letter is to show that in fact the vacuum
functional for
{\it arbitrary} $\bf A$ can be reconstructed from this local expansion
(\ref{eq:YMexp}), given the coefficents $a_i$.

\bigskip
We will work with the functional integral representation
\be
\Psi[{\bf A}]=\int{\cal D}A\,e^{-S[A]-S_b[A,\,{\bf A}]}
\label{eq:funct}
\ee
where $S[A]$ is the Yang-Mills action gauge-fixed in the
Weyl-gauge $A_0=0$. Space-time is Euclidean with co-ordinates
$({\bf x},t)$ and $t\le 0$, so
\be
S[A]=-{\textstyle {1\over g^2}}\int d^3{\bf x}\,dt\,tr\,\left(
{\dot A}^2 + (\nabla \wedge{ A}+{ A}\wedge { A})^2\right)
\ee
The boundary term in
the action is
\be
S_b[A,{\bf A}]=-{\textstyle {2\over g^2}}\int d^3{\bf x}\,tr\,\left(
({\bf A}-A)\cdot{\dot A}\right)|_{t=0}
\ee
The boundary value of $A$ is to be freely integrated over,
i.e. we will not impose a condition such as $A({\bf x}, 0)=0$.
We will assume that at spatial infinity the source $\bf A$ is
a pure gauge ${\bf A}\sim g(\hat{\bf x})^{-1}\nabla g(\hat{\bf x})$.
$S_b$ is chosen so that $\Psi[{\bf A}]$ is invariant under the gauge
transformation $\delta_\omega {\bf A}=\nabla\omega +[{\bf A},\omega]$,
since the effect of varying the source $\bf A$ may be compensated by
gauge transforming $A$.
As $\omega$ cannot depend on time this is
the residual gauge symmetry of $S[A]$ that preserves the gauge
condition.
Functionally differentiating with respect to the source
leads to an insertion of ${\dot A}$.
Given the Wick rotation between Euclidean time, $t$,
and Minkowskian time, this yields the \Sc representation
of the canonical momentum, which in this case is the non-Abelian
electric
field, $E=-ig^2{\delta}/\delta {\bf A}$.

\bigskip
Consider the effect of a scale transformation on the configuration
${\bf A}$ given by
\be
{\bf A}^s({\bf x})={\textstyle{1\over\sqrt s}}{\bf A}(
{\textstyle{1\over\sqrt s}}{\bf x}).
\ee
For small $s$ the scaled field is trivial everywhere except the
vicinity
of the origin, since for ${\bf x}=0$ we have $\A ^s (0)=
s^{-1/2}\A (0)$,
but for ${\bf x}\ne 0\quad \A^s({\bf x})\approx s^{-1/2}g^{-1}
(\hat{\bf x})
\nabla g(\hat{\bf x})$. For large $s$ the field varies slowly in
space
since now $\A^s({\bf x})\approx s^{-1/2}\A(0)$. By studying the
analyticity
of $\Psi [\A ^s]$ in $s$ we will be able to express its value at
$s=1$
in terms of its  value for large $s$, where we can apply the local
expansion
(\ref{eq:YMexp}), and its value for small $s$, which is reliably
computed in
the usual semi-classical perturbation theory and turns out to be
negligible.
More specifically, we will show that $\Psi [\A ^s]$
can be analytically continued to the
complex plane with the negative real axis removed. This enables us
to compute
the contour integral
\be
I(\lambda)={1\over 2\pi i}\int_C {ds\over s-1}\,e^{\lambda (s-1)}
\, \Psi [\A ^s]\label{eq:I}
\ee
in two ways. We take $C$ to be a key-hole shaped contour running
just under the
negative real axis up to
$s=1-s_0$, around the circle of radius $s_0$ centred
on $s=1$ and then back to $s=-\infty$ running just above the
negative
real axis. If we take $s_0$ to be large then we can compute the
integral using
the local expansion for $W$. Each term in this expansion scales
so we can express
it
in terms of $\A$ rather than $\A ^s$,
\bq W[{\bf A}^s]&=&\int d^3{\bf x}\,
(a_1s^{-1/2}tr\,{\bf B}\cdot{\bf B}/m +a_2s^{-3/2}tr\,D
\wedge{\bf B}\cdot
D\wedge{\bf B}/m^3+
\nonumber\\
& &
a_3s^{-3/2}tr\,{\bf B}\cdot
({\bf B}\wedge{\bf B})/m^3 +a_4 s^{-5/2}tr\,{\bf B}\cdot{\bf B}
\,{\bf B}\cdot{\bf B}/m^5+..)
\label{eq:YMexpp}
\eq
This yields an expansion of $\Psi [\A ^s]$ in inverse powers of
$s-1$,
with coefficents that depend on the original configuration,
$\Psi [\A ^s]\sim \sum \,(s-1)^{-n}\psi_n[\A ]$,
enabling us to compute $I(\lambda)$ as
\be
I(\lambda)=\sum_n {\lambda^n \psi_n[\A]\over \Gamma (n+1)}
\ee
We can also evaluate the integral by collapsing the contour
$C$ until it breaks into two disconnected pieces, a small circle
centred on
$s=1$ and a contour that
just surrounds the negative real axis. The integral over the
circle
gives $\Psi [\A]$. By taking $\lambda$
to be real, positive and
very large the contribution from the negative real axis
will be exponentially suppressed, (provided it is not singular,
which we check in perturbation theory), so we obtain for large
$\lambda$
\be
\Psi [\A ]\approx \sum_n {\lambda^n \psi_n[\A]\over \Gamma (n+1)}
\label{eq:gamm}
\ee
which provides a re-summation of the local expansion. Note that only
terms of order up to $s^{-n}$ in (\ref{eq:YMexpp}) will contribute to
$\psi_n$. We might expect to obtain an approximation by truncating
(\ref{eq:gamm}) at some order in $\lambda$.

\bigskip
    As an illustration we will show how the vacuum functional of a
free
massive field theory may be reconstructed from its local expansion.
First scale the
field by setting $\p^s ({\bf x})=s^{-(D-1)/2}\p ({\bf x}/\sqrt s)$.
The vacuum functional
for the scaled field is then $exp\,W [\p^s]=exp-{1\over 2}
\int d^D\x\,
\p\sqrt{-\nabla^2+sm^2}\,\p$ which can be continued to an
analytic function in the
complex $s$-plane with the negative real axis removed, and is
finite at the origin.
For large $s$ we can expand this, or alternatively (\ref{eq:one})
evaluated for $\p^s$, in inverse powers of $s-1$ obtaining the
local series
\be
W[\p^s]=-{m\over 2}\sum_0^\infty {\Gamma (3/2)\over \Gamma (n+1)\,
\Gamma (3/2-n)}
(s-1)^{1/2-n}\int d^D\x\,\p\,\left(1-{\nabla^2\over m^2}\right)^n\p
\ee
which yields an expansion of $\Psi$ in inverse powers of
$s-1$. The term quadratic in $\p$ is just $W[\p^s]$
itself, so if we concentrate on this term rather than the whole
of $\Psi $ we should consider
\be
{1\over 2\pi i}\int_C {ds\over s-1}\,e^{\lambda (s-1)}
\, W [\p ^s] ={m\over 4\sqrt \pi}\sum_0^\infty {(-)^n\lambda^{n-1/2}
\over n!\, (n-1/2)}\int d^D\x\,\p\,\left(1-
{\nabla^2\over m^2}\right)^n\p
\label{eq:sum}
\ee
The integrals will exist for all $n$ provided $\p$ has a momentum
cut-off, $k_0$ say,
but the integral will converge for all $\lambda$ and $k_0$ because
of the
$n!$ in the denominator. This is in contrast to
our original expansion (\ref{eq:one}) which only converges
for $k_0< m$. The series (\ref{eq:sum}) results from expanding in
powers
of $\lambda$ the exponentials in
\be
-{m\over 2\sqrt\pi}\int d^D\x\,\p\,\left({\textstyle 1\over\sqrt
\lambda}
\,{e^{-\lambda (1-{\nabla^2/ m^2})}}+\int_0^\lambda d\lambda
{\textstyle 1\over\sqrt\lambda}
\,{e^{-\lambda(1-{\nabla^2/ m^2})}}\left(1-{\nabla^2\over m^2}\right)
\right)\p ,
\ee
which is just
\be
-{1\over 2}\int d^D\x\,\p\sqrt{-\nabla^2+m^2}\,\p +{m\over
 4\sqrt \pi}
\int d^D\x\,\p\,\left(\int^\infty_\lambda d\lambda\,
{\textstyle 1\over{\sqrt\lambda}^3}
\,{e^{-\lambda (1-{\nabla^2/ m^2})}}\right)\p
\label{eq:sint}
\ee
We see that as $\lambda\rightarrow \infty $ the series
tends to $W[\p]$, as it should. Furthermore, if we keep
$\lambda$ large,
but finite,
the error in approximating $W[\p]$ by the series (\ref{eq:sum}),
as given by the last integral in (\ref{eq:sint}), is
exponentially suppressed. For a finite value of $\lambda$ we can also
truncate the alternating series (\ref{eq:sum}) at order $\lambda^n$
with an error smaller than the first neglected term, so that by taking
$n$ sufficently large in comparison to $\lambda$ and $k_0$
this error can be made small. In conclusion, we can represent $W[\p]$
for cut-off, but otherwise arbitrary, $\p$, by the {\it local} series
(\ref{eq:sum}) truncated at order $\lambda^n$ where $\lambda$ and $n$
are both large and chosen to give acceptable error.


\section{Analyticity of the Vacuum Functional}
To construct an analytic continuation of $\Psi [\A ^s]$ we
adopt a similar approach to that used in \cite{Paul},
complicated by having to work with three spatial dimensions.
We consider scaling each dimension separately, so we set
$\A^{\bf s}=(s_1s_2s_3)^{-1/6}\A (x^1/\sqrt s_1, x^2/\sqrt s_2,
x^3/\sqrt s_3)$ and show that $\Psi [\A ^{\bf s}]$ is analytic
in $s_1,s_2,s_3$ separately.
Firstly
we interchange the names of the Euclidean time, $t$, and one of the
spatial co-ordinates, $x^1$ say, in the functional integral
(\ref{eq:funct})
which we now interpret as the Euclidean time-ordered vacuum
expectation value
\be
\Psi [\A ^{\bf s}]=T\langle 0_r|\,exp\,\left({\textstyle
{2\over g^2}}\int dx^2\,dx^3\,
dt\,tr\,\left( \A^{\bf s}\cdot A^\prime\right)\right)_{x^1=0}\,
|0_r\rangle
\ee
where $|0_r\rangle$ is the vacuum for the Yang-Mills Hamiltonian
defined on the
space $x^1\le 0$ in the axial gauge $A_1=0$ with a boundary
term in the
action ${\textstyle 1\over g^2}\int dx^2\,dx^3\,
dt\,tr\,\left( A\cdot A^\prime\right)$, where the $\prime$ denotes
differentiation
with respect to $x^1$. Expanding the exponential gives
\bq
&&\Psi [\A ^{\bf s}]=
\nn
&&\sum_n\int_{-\infty}^\infty dt_n   \int_{-\infty}^{t_{n-1}}
dt_{n-1}
..\int_{-\infty}^{t_3} dt_2\int_{-\infty}^{t_2} dt_1
\prod_{i=1}^n\left({\textstyle {-{1\over g^2}}}\int dx^2_i\,dx^3_i
\,\A^s_{R_i}
(t_i,x_i^2,x^3_i)\right)
\nn
&&
\langle 0_r|A^\prime _{R_n}(0,x^2_n,x^3_n)\,e^{(t_{n-1}-t_n)H}
..
A^\prime_{R_2}(0,x_2^2,x_3^3)\,
e^{(t_2-t_1)H}\,A^\prime_{R_1}(0,x_1^2,x_1^3)\,|0_r\rangle.
\label{eq:expexp}
\eq
Here $R_i$ stands for both Lie algebra and spatial indices.
The time integrals may be done after Fourier transforming the sources.
To do this we define the $s_1$-independent Fourier mode
\be
a(k,x_2,x_3)\equiv
(s_2s_3)^{-1/6}\int dt\,e^{-ikt}\,\A (t,x^2/\sqrt s_2 ,
x^3/\sqrt s_3),
\ee
so that
\be
\A ^{\bf s} (t,x^2,x^3)={\textstyle {1\over 2\pi}}\int dk \,
e^{ikt/\sqrt s}\,
s^{-1/6}\,a(k,x^2,x^3).
\ee
Substituting this into (\ref{eq:expexp}) gives
\bq
&&\Psi [\A ^{\bf s}]=
\nn
&&\sum_n
\prod_{i=1}^n\left({\textstyle {-{1\over \pi g^2}}}\int
dk_i\, dx^2_i\,dx^3_i\,s_1^{1/3}\, a_{R_i}
(k_i,x_i^2,x^3_i)\right)\,\delta\left( \sum k_i \right)
\nn
&&
\langle 0_r|A^\prime _{R_n}(0,x^2_n,x^3_n)\, {1\over
\sqrt s_1 H-i\sum_1^{n-1}k_j}
..
{1\over \sqrt s_1 H-ik_1}\,A^\prime_{R_1}(0,x_1^2,x_1^3)\,
|0_r\rangle.
\label{eq:expa}
\eq
This makes explicit the $s_1$-dependence of $\Psi [\A^{\bf s}]$.
Although $s_1$ was originally real and positive we may use this
expression to define an analytic continuation to complex values,
yielding
a function that is analytic away from the zeroes of the
denominators in
(\ref{eq:expa}). Since the eigenvalues of the Hermitian
Hamiltonian are real,
the singularities lie on the negative real $s_1$-axis.
Similarly we may show
that $\Psi [\A^{\bf s}]$ continues to an analytic function in
$s_2$ and $s_3$
and, by setting $s_1=s_2=s_3=s$, that $\Psi [\A^{ s}]$ continues to
an analytic function in $s$ on the complex plane with the
negative real axis
removed.



\bigskip
We have seen that for small values of $s$ the configuration
$\A^s$ is non-trivial
only over a short distance about the origin. Since asymptotic
freedom under-writes
semi-classical perturbation theory at short distances we can
reliably use this to
calculate $\Psi [\A ^s]$ for small $s$. We will assume that
Symanzik's work on the
vacuum functional of $\varphi^4$ theory \cite{Sym} can be
generalised to the present
case, and further, that gauge invariance implies that the
source $\A$ needs no
renormalisation
\cite{lusch}. Thus we assume that the cut-off necessary to
define
(\ref{eq:funct}) may be removed leaving a finite vacuum
functional that depends on
the source $\A$, an arbitrary mass-scale $\mu$ and a
renormalised coupling, $g(\mu)$.
(This assumption is supported by the scaleing behaviour
observed
in lattice estimates of $\Psi$, \cite{Jef2}, \cite{Jef3},
\cite{ari}).
Now if in computing $\Psi [\A ^s]$ we were to choose
a new unit of length so as to undo the scaleing of $\A$,
and at the same time
we scale $\mu$ appropriately, then nothing would change
so we must have
that $\Psi [\A^s]$ computed using $\mu$ and $g(\mu)$ is equal to
$\Psi [\A , ]$ computed using  $\mu /\sqrt s$ and
$g (\mu /\sqrt s )$.
As $s$ decreases so does the
coupling, since in perturbation theory $g^{-2}(\mu/\sqrt s)
=g^{-2}(\mu)-11N\,(ln\,s)\,/(48\pi)$ for gauge group $SU(N)$,
so for small enough $s$ it is sufficent to take the
tree-level approximation
\be
\Psi [\A ]\approx e^{-S[A_{cl}]-S_b[A_{cl},\,{\bf A}]}
\ee
where $A_{cl}$ satisfies the Euler-Lagrange equation
$\delta (S[A]-
S_b[A,\,{\bf A}])=0$ under variations of $A$ that are
arbitrary at the
boundary $t=0$.
Now
\bq
&&
\delta\left({\textstyle {1\over 2}}\int d^3{\bf x}\,dt\,tr\,
\left(
{\dot A}^2 + (\nabla \wedge{ A}+{ A}\wedge { A})^2\right)
+\int d^3{\bf x}\,tr\,\left(({\bf A}-A)\cdot{\dot A}\right)
|_{t=0}\right)
=\nn
&&\int d^3{\bf x}\,dt\,tr\,\left((\dot A\,\delta A)^{\dot{}}
-\delta A \left({\mathaccent "7F A} + \nabla\wedge B +A\wedge B+
B\wedge A\right)\right)\nn
&&+\int d^3{\bf x}\,tr\,\left(({\bf A}-A)\cdot{\delta\dot A}
-\delta A\cdot{\dot A}\right)|_{t=0}
\eq
where $B$ is the non-Abelian magnetic field constructed from $A$.
The first and last terms cancel using Stokes' theorem so we
require that
$A_{cl}$ satisfy the Euclidean Yang-Mills equations
${\mathaccent "7F A} +
\nabla\wedge B +A\wedge B+B\wedge A=0$ with boundary condition
$A|_{t=0}=\A$. The boundary integral $S_b[A_{cl},\,{\bf A}]=0$,
so we are
left with an expression for the small-$s$ dependence of
$\Psi [\A ^s]$
\be
\Psi [\A ^s]\approx s^{-{11N\over 48\pi}\int d^3{\bf x}dt
\,tr\,\left({\dot A}_{cl}^2+B_{cl}^2\right)}\equiv s^\alpha
\ee
which, being a positive power of $s$, goes to zero with $s$.
This gives a contribution to $I(\lambda)$ from the cut along
the negative real
axis in the vicinity of the origin of
\be
{1\over 2\pi i}\int {ds\over s-1}\,e^{\lambda (s-1)}\,s^\alpha
={sin\,(\pi\alpha)\over\pi}\,e^{-\lambda}\int_0^\infty
{dx\over 1+x}\,
e^{-\lambda x}x^\alpha
\ee
Since $1/(1+x) \,< 1/x$ for positive $x$, the last integral
is less than
$\lambda^{-\alpha}\Gamma (\alpha)$ and we obtain the bound
\be
|{1\over 2\pi i}\int {ds\over s-1}\,e^{\lambda (s-1)}\,
s^\alpha\,|\,<\,
{|sin (\pi\alpha)\,|\,\Gamma (\alpha)\over \pi\lambda^\alpha
\,e^\lambda}
={1\over |\Gamma (1-\alpha ) \,|\, \lambda^\alpha \,e^\lambda},
\ee
which is negligible for sufficently large $\lambda$.
So we conclude that, for large $\lambda$ we can ignore this
semi-classical contribution to $I(\lambda )$ and reconstruct the
vacuum functional   as
$\Psi [\A ]\approx \sum_n {\lambda^n \psi_n[\A]\over \Gamma (n+1)}$,
where the $\psi_n$ are {\it local} functionals of the field
computable from a knowledge of the vacuum functional evaluated for
fields that vary slowly on the scale of the lightest glueball mass.

\end{document}